# Petascale Cloud Supercomputing for Terapixel Visualization of a Digital Twin

Nicolas S. Holliman, *Member IEEE Computer Society*, Manu Antony, James Charlton, Stephen Dowsland, Philip James and Mark Turner

**Abstract**— Background—Photo-realistic terapixel visualization is computationally intensive and to date there have been no such visualizations of urban digital twins, the few terapixel visualizations that exist have looked towards space rather than earth. Objective—our aims are: creating a *scalable* cloud supercomputer software architecture for visualization; a *photo-realistic terapixel 3D visualization* of urban IoT data supporting daily updates; a *rigorous evaluation* of cloud supercomputing for our application. Method—We migrated the Blender Cycles path tracer to the public cloud within a new software framework designed to scale to petaFLOP performance. Results—we demonstrate we can compute a terapixel visualization in under one hour, the system scaling at 98% efficiency to use 1024 public cloud GPU nodes delivering 14 petaFLOPS. The resulting terapixel image supports interactive browsing of the city and its data at a wide range of sensing scales. Conclusion—The GPU compute resource available in the cloud is greater than anything available on our national supercomputers providing access to globally competitive resources. The direct financial cost of access, compared to procuring and running these systems, was low. The indirect cost, in overcoming teething issues with cloud software development, should reduce significantly over time.

◆

## 1 INTRODUCTION

AS we gather increasing amounts of data about our urban environment it is important to present this in informative, engaging and accessible ways so that the widest possible set of stakeholders have the potential to see the data. The Newcastle Urban Observatory [1] has been collecting IoT sensed environmental data about the city of Newcastle-upon-Tyne for over three years, gathering more than nine hundred million data records to date. As is common to many data platforms the rate of data collection is significantly faster than the rate at which humans can comprehend and learn from the information the data carries [2]. Therefore, we explore how we can present descriptive statistics, such as hourly sensor averages, in a realistic 3D visualization of the city and do so at a range of geographic scales.

Terapixel images are images that contain over one trillion pixels and, within the right toolset [3], provide an in-

---

- *N.S. Holliman is with the School of Computing, Newcastle University, Newcastle-upon-Tyne, NE4 5TG. E-mail: nick.holliman@newcastle.ac.uk*
- *M. Antony is with the School of Computing, Newcastle University, Newcastle-upon-Tyne, NE4 5TG. E-mail: m.antony@newcastle.ac.uk*
- *J. Charlton is with the Department of Architecture and Built Environment, Northumbria University, Newcastle-upon-Tyne, NE1 8ST. E-mail: j.charlton@northumbria.ac.uk*
- *S. Dowsland is with the School of Computing, Newcastle University, Newcastle-upon-Tyne, NE4 5TG. E-mail: stephen.dowsland@newcastle.ac.uk*
- *P. James is with the School of Engineering, Newcastle University, Newcastle-upon-Tyne, NE4 5TG. E-mail: philip.james@newcastle.ac.uk*
- *M. Turner is with the School of Computing, Newcastle University, Newcastle-upon-Tyne, NE4 5TG. E-mail: mark.turner@newcastle.ac.uk*

tuitive, fluid user experience where the viewer can see an overview of the whole image or zoom into incredible detail. In this article we demonstrate that we can zoom in from an overview of just over one square kilometre of the city of Newcastle-upon-Tyne to see detail within a single room in an office or a house with one pixel in the image representing an area of 1.4 mm² in the real world. Because viewing a terapixel image depends only on image display capabilities any web browser can display it, making terapixel images accessible on a wide range of thin clients. This opens access to high quality, high detail visualizations without needing an expensive, in cost or energy use [6], client-side 3D graphics engine. To the best of our knowledge we present here the first terapixel visualization of IoT data within a 3D urban environment.

To visualize the city and its data we have chosen an advanced path-tracing renderer that is more typically used for cinematic and architectural rendering. We selected Cycles, from the Blender toolset [4], because of its high quality physically based lighting simulation calculations. This has allowed us to achieve an elevated level of realism in our rendering of the city and bringing with it graphical options that are not available in visualization tools that use standard hardware rendering libraries.

The combination of high-quality rendering and terapixel imaging can be an attractive one for users and allows us to explore new ways of visualizing urban IoT data within its city context. However, while the end user experience is compelling there is a significant computational cost to producing a high quality terapixel image. To address this issue, we propose the use of supercomputer scale systems in the cloud. The focus of this article is the design, deployment and rigorous evaluation of a scalable cloud rendering architecture for urban data visualization.

## 2 BACKGROUND

We briefly review three background topics: urban data visualization, scalable computing in the cloud, and existing architectures for distributed and cloud rendering.

### 2.1 Urban Data Visualization

Urban data visualization brings together large 3D models of cities and countries with real time and historical data from many sources including IoT sensing devices [5]. The concept of a digital twin for observing, exploring and predicting urban behavior is rapidly gaining ground [6]. Visualization methods directly support the Gemini principle [7] of creating insight into the data in a digital twin.

The Royal Society [8] proposes that for data to be trustworthy it needs to be made intelligently open, in particular, to be accessible, intelligable and assessable. Our investigation of terapixel imaging explores how the accessiblity and intelligibilty of urban IoT data can be improved, potentially leading to better assessability of its reliablity, meaning and value. To achieve this our terapixel 3D urban visualization of Newcastle's IoT data has five goals [9] aiming to be:

*Truthful:* to show accurate statistical data about the city, to scale and situated within its spatial context.

*Functional:* to create an interactive scalable visualization that works across platforms with low overhead.

*Beautiful:* to use the highest quality rendering tools, applying cinematic quality production techniques.

*Insightful:* to reveal new insights across scale about the urban environment in the city of Newcastle-upon-Tyne.

*Enlightening:* accessible to the viewer so they are able to gain a deep comprehension of the data.

Terapixel images have been previously used to visualize astronomical observations [10] while multiple gigapixel images have been used for visualizing cosmology simulation outputs [11]. We aim to apply terapixel imaging to look towards, instead of away from, Earth and to present observations about one region on Earth in detail.

### 2.2 Scalable Cloud Computing

Public cloud services have started to offer as IaaS (Infrastructure as a Service) an increasing amount of GPU compute capacity, this enables supercomputer scale compute performance to be deployed, utilised and released on-demand. Current examples include the Azure N-series [12] and AWS EC2 Elastic GPUs [13].

In the cloud, as with any parallel system, the same classical laws bound a system's scalability. Amdahl's law [14] can be summarized as: any one parallel computation is limited in scalability as the number of nodes $n$ increases by the fraction of the problem that cannot be parallelized. If $p$ is the parallel fraction, and $f$ the serial fraction, then speedup S is given by:

$$S(n) = \frac{1}{f + \dfrac{p}{n}} \tag{1}$$

We need to take care using the cloud that additional management services don't add undue serial overhead and limit our best performance. Even at $f = 0.1\%$ the overhead will limit our best speedup to 506 for $n = 1024$ nodes.

The Gustafson-Barsis' law [15] reconsidered how parallel computers are used in practice and defined scalability in relation to problem size, arguing that the problem size is always scaled to fill the compute capacity:

$$S(n) = n - f(n-1) \tag{2}$$

In the rendering approach we have chosen here the problem can scale to fill the largest machine we have available. As a result, we would expect to see the Gustafson-Barsis law holding if the problem size remains large enough. We should therefore see efficient linear scaling as we add more compute nodes. Even so there is a limit to the compute size of our problem and at that point Amdhal's law will ultimately define the bounds of scalable performance for us.

In later work Gustafson [16] argued the for use of *ends-based* performance measures as well as pure computational metrics. As our interest here is application driven as much as system driven we also consider the following from Gustafson's proposed ends-based metrics:

- Time to compute the answer.
- Completeness of the answer.
- Maximum feasible problem size.
- System reliability.
- Performance divided by the system cost.

We also add an additional application metric, *future scaling*, which provides a prediction of how far into the future it will be before the system performance we achieve on a supercomputer today might be available on everyday desktop workstations.

### 2.3 High Performance Visualization Systems

Many distributed and parallel rendering system architectures for High Performance Visualization (HPV) have been proposed as new High Performance (HPC) and High Throughput (HTC) platforms have appeared [2].

From the earliest days of HPV systems, rendering has often been thought of as ideally parallel even though this is not the case when geometry data sizes exceed local memory or network transfer bandwidth [17]. Nonetheless, rendering is often a good test of new hardware performance since it is capable of scaling to follow Gustafson's law, absorbing all available compute resource. This is particularly the case for photorealistic and physically-based path tracing algorithms whose performance is dominated by floating point calculations.

Remote HPV systems can be categorised as being *send-data*, *send-geometry* or *send-image* [2]. Send-data systems assume a local GPU at the client and the server sends a subset of the application data to be converted to geometry and rendered locally into an image. Send-geometry systems convert the data to geometry at the server and send just geometric and material data to the client to be rendered. In both cases the client needs substantial network, and compute resources, and in the worst case as the data size grows the client-side resource demand is unbounded.

The third approach send-image, and the one we take in this project, is where the geometry is generated and rendered on the server and only the image is sent to the client. The key advantage is the maximum image data required at any one time is bounded by the fixed number of pixels on display rather than the unbounded data or model geometry. This bounds the network bandwidth needed between server and client.

A similar approach to ours was described by Chen et al in [18] which generalises previous image-based rendering systems such as QuickTime VR. Chen uses a multiresolution tiled image as the output format and serves this from an http server to viewers running in client browsers. Our work differs in that we aim to use the public cloud to render the terapixel images and to parallelize all the operations needed to output the multiresolution tiled image format in addition to the rendering operations.

One solution to our visualization problem could be to use a render farm, both private and public render farms exist and render services are available in the cloud. However, currently these don't offer all the functionality we need to build an end-to-end terapixel solution, they are primarily tuned to production animation rendering.

## 2.4 Summary

We aim to generate high quality, terapixel urban data visualizations that are rendered by a system that scales to efficiently use as many compute nodes as are available. We have chosen to design this as a send-image system architecture so that thin-clients are as capable of displaying the visualization as fat-clients with the aim of supporting wide accessibility to the results of advanced visualization methods.

## 3 SYSTEM ARCHITECTURE

We have previously discussed [19] how a send-image architecture for remote cloud rendering, where the rendering is server-based and a pixel-stream is sent to the client, has important advantages including 1) Decoupling the task of display from the computationally complex task of rendering, enabling the use of thin-clients to view complex visualizations, 2) Limiting the maximum bandwidth needed to the client display device, which is otherwise unbounded in thick-client solutions. The following details the requirements and the design of such a system for terapixel visualization.

## 3.1 The Visualization

As shown in Fig. 1 we used a 1.28km x 1.28km 3D map of central Newcastle-upon-Tyne as the context for our visualization. This defined the geographic boundaries of the visualization and the IoT sensors we included. Since a ter-apixel image consists of $2^{20} * 2^{20}$ or $(1,048,576)^2$ pixels this means that one pixel in our terapixel image represents approximately 1.2mm x 1.2mm in the real world. This provides the ability in a single image to zoom in from city scale to an area of about the size of desk in a room in a building. This represents a zoom factor of approximately 512x on a full HD display allowing users to view data in the city at a wide range of scales.

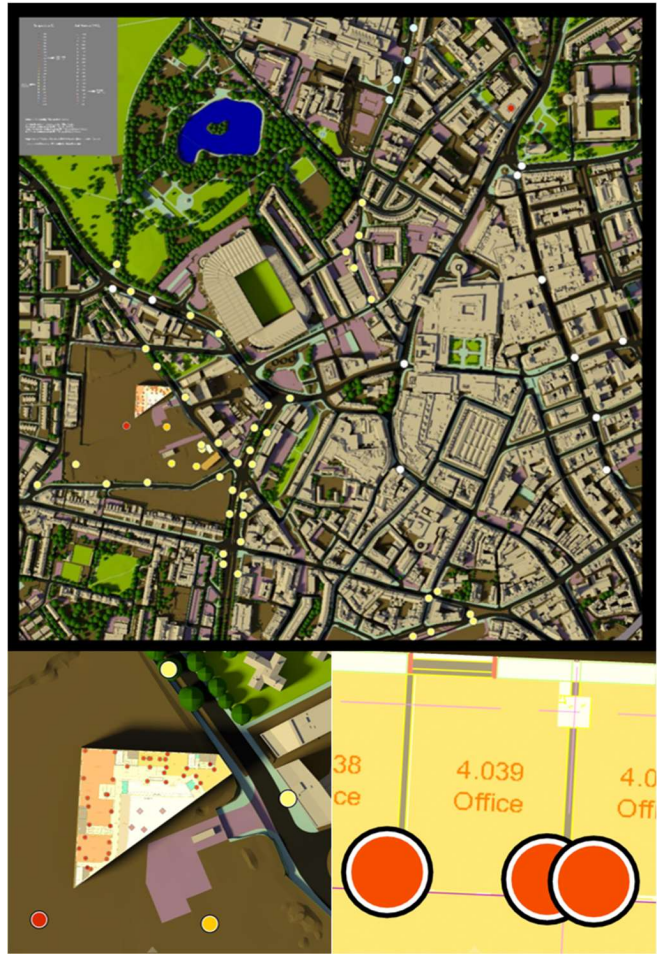

Fig. 1. The terascope image of Newcastle-upon-Tyne, the inset images are zoomed in views illustrating the internal sensors in the Urban Sciences Building, glyph colours represent the average temperature at each sensor over one hour, glyphs are rescaled as the viewer zooms in so that they retain a reasonable size in the image.

Set within this map are glyphs representing the location of Urban Observatory IoT sensors [1] around the city, the sensors are situated both externally in the city and internally within the Urban Sciences Building. For the purposes of this visualization we use an arbitrary day's temperature values, averaged over one hour for each of the sensors. The average temperature value is represented as a solid colour at the centre of each glyph, though not obvious in a single still image the glyphs automatically orient themselves to face the camera and rescale at preset zoom levels to maintain a reasonable size. Some of the glyphs represent sensors that are physically as close as 100mm in the real world.

TABLE 1
TERAPIXEL IMAGE PROPERTIES

| Pyramid level | Image side length (pixels) | Total number of pixels at this level | Number of 512x512 tiles | Tile side length in the real world (mm) |
|---|---|---|---|---|
| **12** | **1048576** | **1099511627776** | **4194304** | **625** |
| 11 | 524288 | 274877906944 | 1048576 | 1250 |
| 10 | 262144 | 68719476736 | 262144 | 2500 |
| 9 | 131072 | 17179869184 | 65536 | 5000 |
| **8** | **65536** | **4294967296** | **16384** | **10000** |
| 7 | 32768 | 1073741824 | 4096 | 20000 |
| 6 | 16384 | 268435456 | 1024 | 40000 |
| 5 | 8192 | 67108864 | 256 | 80000 |
| **4** | **4096** | **16777216** | **64** | **160000** |
| 3 | 2048 | 4194304 | 16 | 320000 |
| 2 | 1024 | 1048576 | 4 | 640000 |
| 1 | 512 | 262144 | 1 | 1280000 |

The terapixel image is a hierarchical pyramid of 512x512 pixel tiles, this supports interactive panning and zooming for the image.

## 3.2 Technical Requirements

One terapixel image requires approximately one terabyte of storage, something that is not currently easy to download as a single image file to most clients. Instead, we manage image viewing as a streaming process using a client-side image streaming tool, the krpano Viewer [3]. This requires the image to be stored on the server in a hierarchy, with each layer of the pyramid a complete image, tiled into 512x512 image tiles. The properties of each level in the pyramid are shown in Table 1 where the side length of each layer doubles with depth in the pyramid.

To create the image pyramid, we could render each level individually or render just the highest resolutionlevel 12 and sub-sample this to calculate all the lower levels in the hierarchy. However, we adopt a compromise where we render levels 12, 8 and 4 and subsample these levels to fill in each set of three intermediate levels. This saves total rendering time compared to rendering every level and reduces compression errors in the lower level images compared to repeatedly subsampling the full image. This approach also allows us to render features such as the IoT sensor glyphs at different scales in the image for the three key hierarchy levels, keeping the glyphs a more appropriate size as the viewer scales through the wide zoom range in the image.

To calculate intermediate levels in the image pyramid from rendered levels we can apply a general smoothing function [20]. Any layer *l-1* can be generated from a higher resolution level *l* by sub-sampling using a kernel weighted filter such that a sub-sampled pixel $G_{l-1}(i,j)$ is related to the pixels in a higher resolution image by:

$$G_{l-1}(i, j) = \sum_m \sum_n w(m, n) G_l(2i + m, 2j + n) \quad (3)$$

where the weighting function *w(m,n)* is chosen to balance between the speed of computation and quality of the sub-sampled layers.

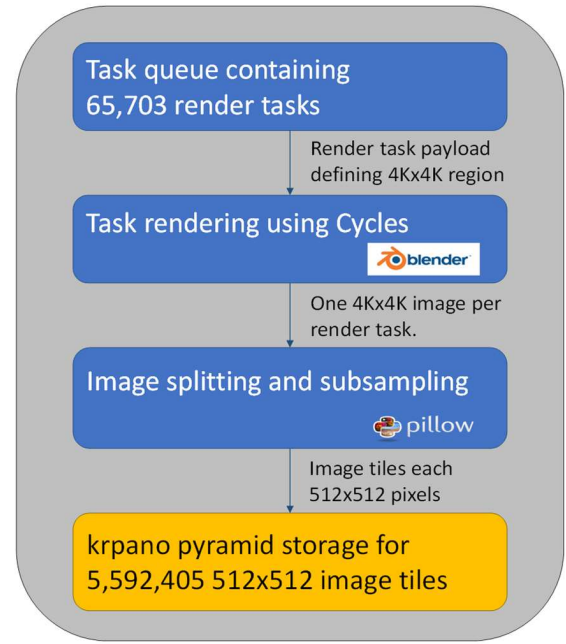

Fig. 2. The data flow for the compute tasks required to generate a single terapixel image pyramid. Blender is used to render the images and Pillow is used to generate the krpano image tiles.

## 3.3 Design

The computation we need to undertake is now defined. To generate a one terapixel image we need to compute an image pyramid of 5,592,405 image tiles. Every fourth level in the pyramid will be rendered directly and the remaining intermediate levels will be computed by subsampling the rendered layers.

As shown in Fig. 2 we group image tiles for rendering efficiency render into tasks of 4096x4096 pixels which are then split and subsampled as required into tiles. We therefore need to compute a total of 65,703 render tasks. For the examples here, our entire 3D city model is assumed to be replicable to every compute node while each task definition has a relatively small payload defining the 4Kx4K region to be rendered. The result is an almost ideally task parallel application, with little communication overhead compared to compute demand. This makes it suitable as a starting point for testing the scalability of cloud supercomputing.

## 4 SYSTEM IMPLEMENTATION

In order to implement a highly scalable parallel version of the data flow architecture in Fig. 2 we used a combination of Microsoft Azure tools and our own application software.

### 4.1 Scalable public cloud framework

We previously implemented a rendering framework for urban IoT data that was designed to update images from live IoT data using a mixture of public and private cloud resources [21]. This was capable of rendering images up to UHD (3840x1920) resolution and was used to update images of IoT data every few minutes throughout the day.

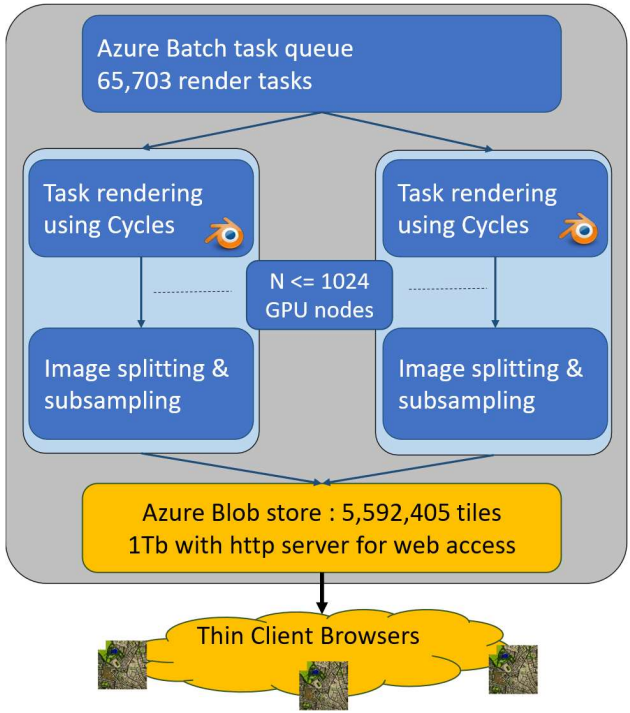

Fig. 3. Cloud supercomputing system architecture, constructed over Azure™ public cloud IaaS.

The size of the computational task for the terapixel image was too large for our private cloud. We therefore created the terapixel system using public cloud components. To create a scalable frame work for the computation we used the following components as shown in Fig. 3:

- We inject the 65,793 rendering tasks into an Azure Batch task queue and allow the queue to distribute the tasks between the pool of compute nodes. It also manages node failure and replacement. The tasks are added to the queue using Azure Functions, a server-less service that runs single scripts on-demand. The script creates three separate jobs, one each for levels 4, 8 and 12 in the image pyramid, and then creates the render tasks for each job. Each task contains the data required to start Blender when run on a node.
- Compute nodes are packaged as an Azure Managed Image that runs the Blender and Pillow tools with the whole city model. A pool of compute nodes is then requested manually, using the managed image as a template. The number of compute nodes is varied to adjust for the performance data collection. Note that compute nodes can, and do, fail during execution and Azure should seamlessly swap these back in.
- Azure Blob Storage provides a scalable solution to storing the 5,592,405 output image tiles in a folder hierarchy suitable for the krpano viewing tool. By default Blob Storage has an http server interface suitable for serving the client-side krpano app with the image tiles it requests when a user is viewing the image.

While much of the system underpinnings are managed by Azure we still had to run custom health checks on the nodes to ensure that drivers versions were correct and that the GPU was still accessible to the Blender. On some occasions we found that driver issues arose that required action to update and reboot the machine. Healthchecks and required actions were carried out using Ansible [22]. Ansible is an IT automation engine, which allowed us to run arbitrary commands across our batch pool through an SSH connection. Ansible requires the creation of an inventory in order to connect to a pool of machines. This inventory was generated using the Azure Batch API [23].

### 4.2 Blender Cycles Rendering

Blender Cycles is an unbiased physically-based path tracing engine which creates an image by tracing light rays backwards from the camera through a scene. Because it simulates realistic effects such as soft shadows, caustics and glossy surfaces it can require very significant compute times. However, also because it supports these effects it allows the use of advanced cinematic visualization effects in an urban smart city visualization.

There are a considerable number of optimisation and tuning parameters in Blender Cycles, and we ran a series of trials with the aim being to optimize image quality and compute time. In particular for GPU supported execution we set the number of samples per pixel to be 20, the Cycles tile size to be 256x256 and the post-rendering image

TABLE 2
TERAPIXEL TASK PROPERTIES

| Hierarchy level | 4kx4k pixel image rendering tasks | 512x512 pixel output tiles | Estimated storage for JPEG coded tiles in kB |
|---|---|---|---|
| **12** | 65536 | **4194304** | **436207616** |
| *11* | | *1048576* | **109051904** |
| *10* | | *262144* | **27262976** |
| *9* | | *65536* | **6815744** |
| | | | |
| **8** | 256 | **16384** | **1703936** |
| *7* | | *4096* | **425984** |
| *6* | | *1024* | **106496** |
| *5* | | *256* | **26624** |
| **4** | 1 | **64** | **6656** |
| *3* | | *16* | **1664** |
| *2* | | *4* | **416** |
| *1* | | *1* | *104* |
| **Totals** | **65,793** | **5,592,405** | **581,610,120** |

One terapixel image requires 65,793 task computations generating 5,592,405 image tiles in 12 hierarchical levels requiring approximately 0.54 tB of Azure Blob™ storage, levels 4, 8 and 12 are computed, other levels are downsampled from these results denoising radius to be 5. This resulted in a single 4Kx4K task rendering time of approximately 150 seconds on the target hardware described below.

### 4.3 Pillow for Image Splitting and Sub-sampling

The output from Blender for each render task is a 4kx4k image and this needs to be split into 512x512 tiles and subsampled to form the tiles for the three lower resolution layers beneath each rendered layer in the pyramid, as described in Table 2.

To retain image quality for the splitting process Blender is requested to output a lossless PNG format image. This is then passed to a python script calling the Pillow API which both splits the 4Kx4K image into 512x512 tiles and subsamples it to form the lower resolution layers using the resize function. All 512x512 tiles are stored using JPG to save storage space and client transfer bandwidth.

### 4.4 Performance metric collection

To understand the performance of the system we designed into the architecture a series of performance metric data collection points. These generate typed, timed messages at key points in the computation, for example rendering start and end points, that are stored in a dedi-

TABLE 3
TERASCOPE PERFORMANCE METRICS

| Metric | Units | Frequency |
|---|---|---|
| GPU Utilization | percentage | 1 Hz |
| GPU Temperature | degrees Centigrade | 1 Hz |
| GPU Power | Watts | 1 Hz |
| Render duration | seconds | per task |
| Tiling duration | seconds | per task |
| Storage duration | seconds | per task |

During each run of the system we collect a continuous stream of data from all the operational compute nodes in addition to Azure's standard metrics, a subset is shown here.

cated database. Table 3 summarises the subset of metrics we collect that are most relevant to this article.

The performance capture system is built on software commonly refered to as the ELK stack. ELK being a combination of Elasticsearch, Logstash and Kibana. Elasticsearch is a distributed, RESTful search and analytics engine [24]. Logstash is a data processing pipeline that ingests data from multiple sources, transforms it, and sends it on to a storage system. Kibana is a visualization tool for exploring data stored in Elasticsearch.

Data stored in Elasticsearch is arranged into indexes. Indexes are collections of data points that share the same data model. The Terascope performance metrics use three different indexes; *render*, *gpu* and *metricbeat*. All the data stored in each of these indexes is timeseries in nature – the value of metric X at time Y was Z. The render index stores all the render events as an image is generated on the node. The GPU index stores all the performance metrics that can be interogated from the GPU hardware for every node in the entire pool. GPU data is captured by logstash every second and sent to the Elasticsearch GPU index. Finally, the metricbeat index is populated using a logstash tool called metricbeat. Metricbeat captures data about the state of all the hardware (with the exception of the GPU) on each of the nodes. There are many granular details to this data but broadly it captures information about the state of the CPU cores, memory, disks and network.

These metrics complement the metrics generated by Azure which don't record at the same level of granularity.

This approach also allowed us to compare our own measures of run time with Azure's measures providing a way to track any overhead introduced by the Azure framework.

The total runtime system load for our metrics collection is not significant compared to the overall compute load and each full terapixel image run generates around 65 Gigabytes of performance data.

## 5 EVALUATION

### 5.1 Characterising the tasks

The computation of a terapixel image using Blender Cycles should be ideally parallel since we can, in the current scene replicate, the model geomery on every node. The main communications load is the output to Azure Blob storage of a 4Kx4K image divided into 512x512 tiles every task, roughly every 150 seconds. A characterisation of the tasks we need to compute is in Table 3.

### 5.2 Choosing a cloud virtual machine

Current public cloud offerings present a range of predefined virtual machine specifications to choose from and a number of possible global locations. For the compute nodes in the first set of evaluation runs reported here we choose a lower end virtual machine to control our costs, the Azure NC6 node. This consists of an Intel Xeon E5-2690v3 CPU, a K80 NVIDIA GPU with 56Gbytes of memory and 380Gbytes of SSD disk space. Each NC6 virtual machine however shares the CPU and the GPU with a second virtual machine and has half the CPU cores and one half of the K80 GPU card. This results in each NC6 node having six CPU cores and one GPU. Therefore, the theoretical GPU compute performance of each NC6 node is approximately 2.8TFLOPS in single precision mode [25] which is the mode used by Blender. The largest pool size we requested was 128 GPU NC6 nodes, this provided a theoretical peak performance of 0.36 PFLOPS.

In addition, we use Azure resources to organize the computation: one node for running the task queue and one terabyte of Azure blob storage for the image tiles making up the terapixel image. An additional node hosts an Elasticsearch database that is used for storing all the performance metric data generated during each run.

### 5.3 System scaling pilots

To pilot the scaling performance of the system we undertook two evaluations, an initial series of runs generating a gigapixel image (32678x32768 pixels) and a second computing the full terapixel image (1048576x1048576 pixels).

The gigapixel test was used to evaluate scaling over the full range of compute nodes from 1 to 128 nodes. However, as the gigapixel image consists of just 256 render tasks we predicted a drop off in scaling efficiency with higher numbers of compute nodes. This was because the variation in run time between tasks means the number of tasks available per node could drop to less than two for some nodes and as a result at the end of the computation some nodes are left idle for significant amounts of time.

The gigapixel results are summarised in Fig. 4. The system scales reasonably, although sub-linearly, up to 96 nodes. Then the scaling efficiency reduces markedly as predicted, down to 63% of expected speedup at 112 and 128 nodes.

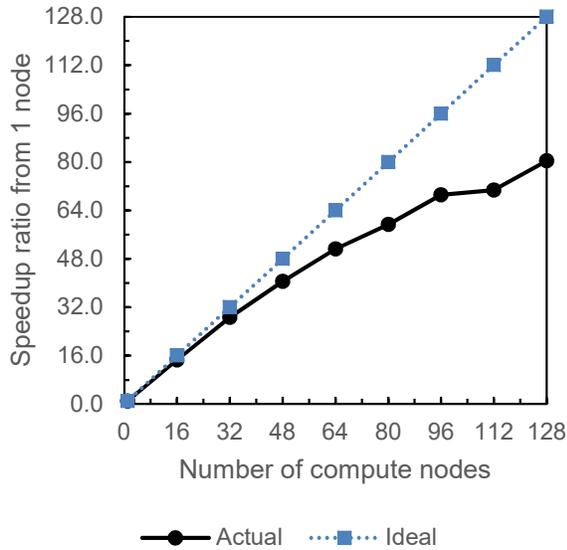

Fig. 4. Evaluating the scalability of the gigapixel computation on NC6 K80 nodes, speedup as a function of the number GPU nodes from 1 to 128 nodes scales, as expected, sub-linearly due to low task load.

We expected noticeable variation in results from run to run due to load imbalances in the computation, to resource contention for physical hardware or to nodes failing and being automatically replaced by Azure. We tested this variability using repeated runs and saw run times that varied by up to +/-8.5% from the mean.

The gigapixel pilot results are broadly as expected with the system performing well up to the point where the amount of work available is no longer enough to keep each node fully busy. It therefore failed the Gustafson-Barsis' scaling law as the computational demand was not a good match to the compute resources available. With this pilot of the system architecture complete we moved on to evaluate the full terapixel computation.

The terapixel experiment was more challenging, requiring substantially more compute time and output storage. For these reasons the smallest number of nodes we evaluated the terapixel computation with was 64, and even with 64 GPUs the total run time was approximately 45 hours. In contrast to the gigapixel image we predicted the results from the terapixel experiment should exhibit close to linear speedup as the number of compute nodes scales from 64 to 128 nodes because the amount of work in the terapixel computation easily exceeds the quantity of compute resource available.

The measured scaling results for the terapixel image are shown in Fig. 5. These appear to show that the system is scaling better than linearly, i.e. as more compute nodes are added the system is running faster than we would predict.

Investigating this result in detail we found a small number of outlier tasks had affected some runs. These outliers ran for 70 times longer on average than we would

| Run | Time lost to outlier tasks (s) | Lost nodes (equivalent) | Normalised run time (s) | Normalised node count |
|---|---|---|---|---|
| 64 | 159305 | 1.00 | 150550 | 63.00 |
| 80 | 49596 | 0.40 | 119815 | 79.60 |
| 96 | 27581 | 0.27 | 98909 | 95.73 |
| 112 | 60151 | 0.75 | 81964 | 111.25 |
| 128 | 0 | 0.00 | 73268 | 128.00 |

Outlier tasks were a small percentage of total tasks that ran without GPU acceleration these caused an inflated measured speedup ratio to be calculated, normalized run time and node count removed these outlier tasks from consideration.

have expected. Further investigation of the task performance metrics uncovered that a small number of rendering tasks were running without GPU acceleration in Blender. This was not a consistent proportion on every run and the effect is summarised in Table 4.

As a result of the outliers we needed a method to estimate speedup fairly, to do this we excluded outlier tasks from consideration in the scaling calculations. The effect of the outlier tasks is removed from both from the run times of nodes and from the available node GPU resource. The resulting normalised speedup is also shown in Fig. 5.

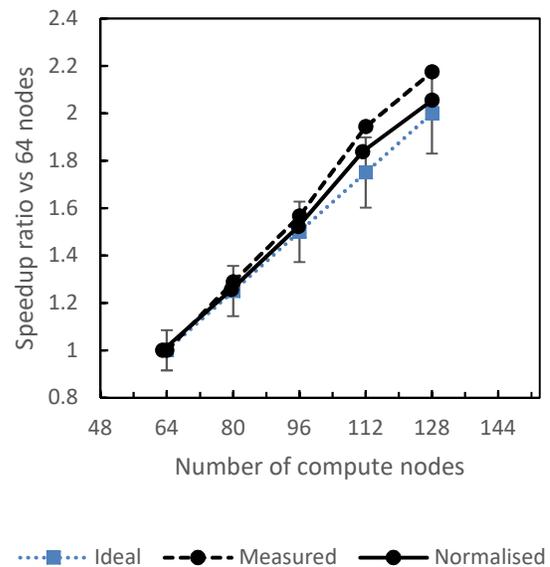

Fig. 5. Evaluating the scalability of the terapixel computation on NC6 K80 nodes, plotting speedup ratio as a function of the number of GPU nodes from 64 to 128 nodes, the bars show the +/-8.5% range of variability we measured in repeated runs.

This shows the speedup ratio improving close to linearly as expected, with a small variation from the ideal that is now within the measured +/-8.5% range. In later runs we ran a script to detect and restart nodes that had long running tasks, and subsequently found a versioning issue with drivers on some nodes that we corrected at system initialization before runs began.

We found that our own metrics and those comparable metrics that were collected by Azure were in close agreement regarding run time. However, our metrics collected finer grained results from within the computation in each task allowing verification of the sub-tasks shown Fig. 2.

## 5.4 Scaling to supercomputer performance

Following the two successful pilots we were interested to investigate the scale of computing performance the cloud could deliver for our application. Simultaneously a new GPU node became available in the Azure public cloud and over 1024 of them were available on demand. The new NC6v3 has six cores and one NVIDIA TESLA V100 GPU [26] delivering 14 teraFLOPS of single precision compute. Theoretically this would allow a system of 1024 nodes to scale to over 14 petaFLOPS compute performance, a level equivalent to systems high in the top 500 list of global supercomputers [27].

Based on a series of trial runs we found our single node used for performance metric capture could not cope with the rate of metrics data being generated in these larger runs. We adjusted the architecture so that parallel data stores were used for streaming metrics data. Because of the size of terapixel image computation we started our collection of metrics at 64 nodes and scaled up to collect data for runs at 128, 256, 512, 768 and 1024 nodes.

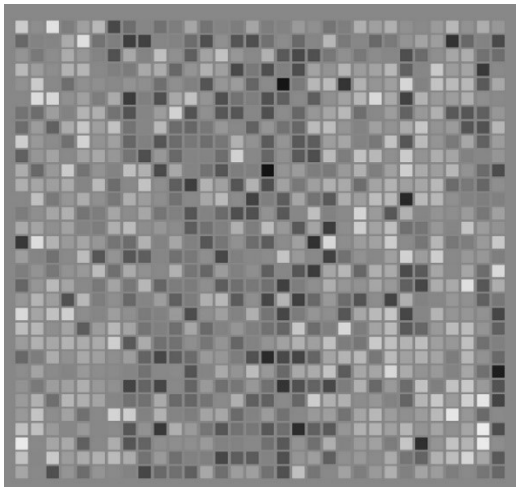

Fig. 7. Heat map of total computation time for all tasks computed by each GPU node in the 1024 run, median is 50% grey, white is nodes executing for more time and black nodes for less time, we expect the random distribution of totals shown as individual task times vary.

Recalling issues in the pilot runs with slow tasks not running on the GPU we confirmed runs were successful by analysing the spread of task run times and the sum of compute time for all tasks on each node. This time there were no outlier tasks in any of the runs. The total compute time per node is shown in Fig. 7 as a heatmap for the 1024 run. As expected, this shows a random spread of total compute time resulting from the natural variations in individual task compute times. However, the results for the 768 run revealed that four nodes had significantly lower total compute values. Analysing total compute time per node across all runs showed the same behaviour in run 256. This turned out to be a standard behaviour where Azure can deallocate and then restore a small number of nodes, in both cases in groups of four. The time taken for this reallocation was approximately twenty minutes of real

TABLE 5
TERAPIXEL NORMALISED PERFORMANCE DATA (V100)

| Run | Run Time (s) | Lost Time (s) | Lost Nodes | Normalised Nodes |
|-----|---------|---------|-------|------------|
| 64 | 45632 | 0 | 0.000 | 64.0 |
| 80 | 36567 | 0 | 0.000 | 80.0 |
| 96 | 30900 | 0 | 0.000 | 96.0 |
| 112 | 26551 | 0 | 0.000 | 112.0 |
| 128 | 22778 | 0 | 0.000 | 128.0 |
| 256 | 11469 | 5467 | 0.477 | 255.5 |
| 512 | 5726 | 0 | 0.000 | 512.0 |
| 768 | 3824 | 5896 | 1.542 | 766.5 |
| 1024 | 2901 | 0 | 0.000 | 1024.0 |

The final column is the normalised number of nodes after accounting for dropped and reinstated nodes in runs 256 and 768, this repeats the pilot runs from 64 to 128 using V100 nodes and then scales up to the maximum tested 1024 nodes.

time, hence those nodes showed significantly lower compute time totals compared to their peers. In order to account for this behaviour we calculated the normalised number of nodes for the scaling calculations, based on the missing compute time, as shown in Table 5.

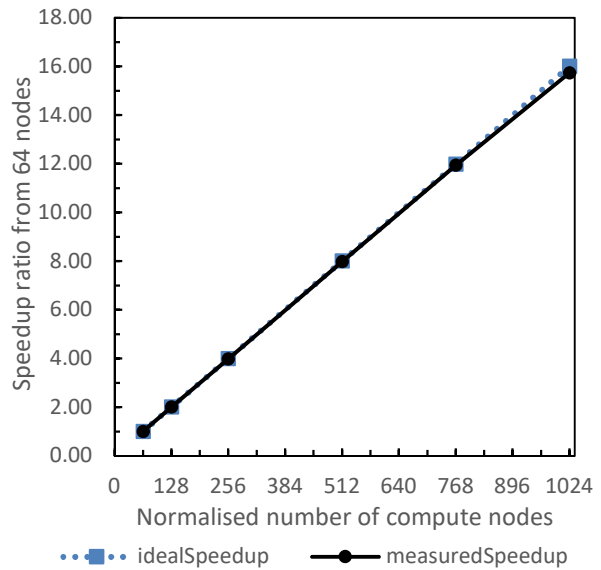

Fig. 8. The speedup ratio plotted for the terapixel computation from 64 to 1024 NC6v3 V100 GPU nodes, the speedup shows close to linear scaling achieving 98% efficiency at 1024 nodes.

Using the normalised number of nodes for each run we are able to plot the scaling graph in Fig. 8. The shows the terapixel calculation scales to make good use of from 64 to 1024 nodes with a calculated efficiency of 98%, utilizing 14 of the maximum 14.3 petaFLOPS available.

An important ends-based performance measure is the *wall clock time* to compute an answer [16]. The total compute time for all tasks in one terapixel image varies slightly from run to run. To estimate the wall clock time to run the terapixel image on a single NC6v3 node we averaged across the total task compute time for each run, resulting in an estimate of 2,940,581 seconds or approximately 34 days of compute time. This compares to 45,632 seconds or 12.7 hours on 64 nodes and to 2,901 seconds or 48 minutes on 1024 nodes, as reported in Table 5.

## 5.5 Accessible visualization platform

One important outcome for us was that the terapixel visualization was accessible to users on thin client devices. The aim being to make results from our supercomputer visualizations open to a range of audiences. Serving the hierarchical image format from Azure cloud Blob store to a krpano viewer enabled web page allowed the terapixel image to be browsed on devices including 24 Mpixel video walls, desktop and laptop computers, UHD TVs, tablets, cell phones and even a gym cycling machine, as illustrated in Fig. 9. You can try it on your own device at this link [28].

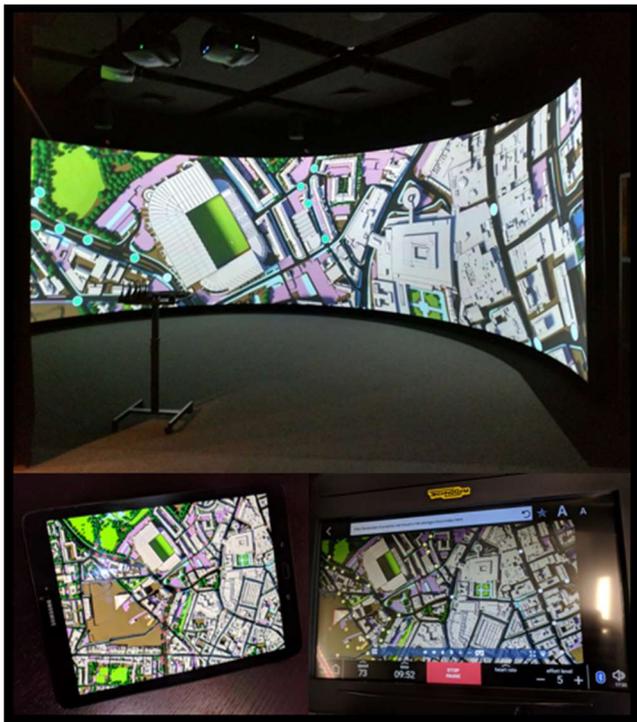

Fig. 9. The hierachical krpano image format for serving the visualization to users from Azure Blob store allows it to be viewed interactively on a wide range of thin client devices, shown here anticlockwise from top are the 8 Mpixel Curtin HIVE cylinder, a Samsung Galaxy S3 and on our local gym cycling machine.

## 6 DISCUSSION

## 6.1 Computational performance

The system, as reported above, did scale to use the resources available, it was not ultimately limited by Amdhal's law and was able to take advantage of Gustafson-Barsis' law application scaling to use the all the resources available efficiently.

However, this success did require software engineering resources to develop and test the system. A number of unforeseen hurdles had to be overcome with specialist help relating to remote systems issues that were difficult to access or difficult to resolve without cloud supplier intervention. These are likely to be teething issues with new cloud systems and compared to the procurement and systems support effort needed to build our own equivalent system were a low barrier to making progress.

The send-image system architecture was successful in allowing us to deliver the resulting visualization to thin-clients across the whole spectrum of display types available to end-users. Accessibility of results to a wide range of stakeholders is an important goal for urban data systems.

## 6.2 Application performance

For end users it is application rather than computational performance that is of paramount importance. We consider Gustafon's [16] ends-based metrics below:

The *time to compute an answer* dropped from an infeasible 34 days to a reasonable 48 minutes on the largest cloud system tested. This fulfills our original end user target of producing a terapixel image once a day.

The degree *of completeness of the answer* is more complex to assess as in photo-realistic graphics we can always do more. However, we have produced the first terapixel visualization of an urban digital twin and done so with high quality 3D geometry, rendered using path tracing with a good sampling level of up to 20 rays per pixel, or a maximum of 20 trillion rays cast.

The *maximum feasible problem size* we measure by image resolution in pixels, the cloud enabled the production of a terapixel image that would be impractical on a desktop system. This is also the highest resolution we need for the current 1.2kmx1.2km city model as we don't have urban sensors or city geometry at sub-millimeter scale.

*System reliability* is critical in long or large rendering operations. The self-repairing behavior demonstrated by the cloud, even though more complex to account for in scaling calculations, is a benefit. It is something that we did not have in our private cloud [21] where days of runtime could be lost while system repairs took place. This depends on local investment in system support.

*Hardware updates* from an application viewpoint the change in the cloud from NVIDIA K80 to V100 GPU nodes was transparent delivering about three times improvement in performance. This would have been a very significant procurement effort for an upgrade to a local supercomputer installation.

*Confidential data protection* building a send-image architecture in the cloud means confidential data, particularly the 3D model of the city was protected. This helped enforce license restrictions as no model data was sent to

### TABLE 6
#### TERASCOPE SYSTEM ENERGY USE

| $n$ nodes used | $p_{av}$ average power draw (kW) | $r_{norm}$ normalised run time (hrs) | $E$ total power (kWh) |
|---|---|---|---|
| 64 | 7.44 | 41.82 | 311 |
| 80 | 9.47 | 33.28 | 315 |
| 96 | 11.42 | 27.47 | 314 |
| 112 | 13.66 | 22.77 | 311 |
| 128 | 15.02 | 20.35 | 306 |

Energy use summarised per run on the NC6 K80 nodes, more nodes use more energy but run for a shorter time so total power drawn is similar across runs.

client devices, this is in contrast to webGL browser solutions where model data is sent to client devices.

### 6.4 Energy use and financial cost

On-premises supercomputers require long term commitments to energy and running costs, whereas for cloud supercomputers we can consider these as one-off costs for each run. For the terapixel runs we predict that both energy use and costs should be constant with number of nodes as the compute time falls linearly with the number of nodes used, as shown in Fig. 6.

We estimate total energy per run, $E$, in kWh as

$$E = p_{av} * r_{norm} \qquad (3)$$

where $p_{av}$ is the average power drawn in kW by the system and, $r_{norm}$, is the normalized total run time on $n$ nodes. The results are shown in Table 6 were as expected $E$ remains approximately constant across runs.

To understand our experimental costs, we similarly estimated total cost per run, $C$, as

### TABLE 7
#### TERASCOPE SYSTEM COSTS

| $n$ Nodes used | $c_{hr}$ cost per hour (£) | $r_{norm}$ Normalised run time (hrs) | $C$ Total Cost (£) | $PP$ Pixels/Pound (millions) |
|---|---|---|---|---|
| 64 | 42.94 | 41.82 | 1796 | 612 |
| 80 | 53.68 | 33.28 | 1787 | 615 |
| 96 | 64.42 | 27.47 | 1770 | 621 |
| 112 | 75.15 | 22.77 | 1711 | 642 |
| 128 | 85.89 | 20.35 | 1748 | 629 |

Costs are summarized for each run on the NC6 K80 nodes, the costs remain similar across runs because the total compute time required for the terapixel computation is constant.

$$C = c_{hr} * r_{norm} \qquad (4)$$

where $c_{hr}$ is the average cost per hour of the number of nodes. Table 7 shows costs stay roughly constant as expected. It is worth noting this doesn't include development time or failed runs, we estimated these factors doubled our costs for this set of experiments, but in a longer series of production runs would become less significant.

A final ends-based metric from Gustfson is *performance divided by system cost*, which we estimate for each run as the ratio of pixels per pound spent, $PP$.

$$PP = \frac{Pix}{C} \qquad (5)$$

where $Pix$ is one trillion, the total number of pixels in our rendered image. If our scaling is sub-linear, we will see this ratio go down and see diminishing returns on our investment. Because in the terapixel case our scaling is close to linear $PP$ is approximately constant varying by less than 5%, as shown in Table 7.

### 6.3 Future scaling using the cloud

We have demonstrated the ability to scale an application on current generation GPU technology using the cloud to over 1000x the performance of a single desktop system. As application researchers this *future scaling* allows us to experiment with creating the software tools that we need for the next generation of visualization applications. How far into the future this allows us to plan depends on predicting how well system performance will improve.

The future of Moore's law is being challenged and scaling out to the cloud has been suggested as one way to mitigate the need for continuous hardware improvement. However, while we have demonstrated the cloud can provide huge advantage during any single hardware cycle it, at best, scales up linearly. Over time this could not match the exponential scaling Moore's law has delivered. Even if datacenters could scale exponentially by floor area so would the power and cooling needs.

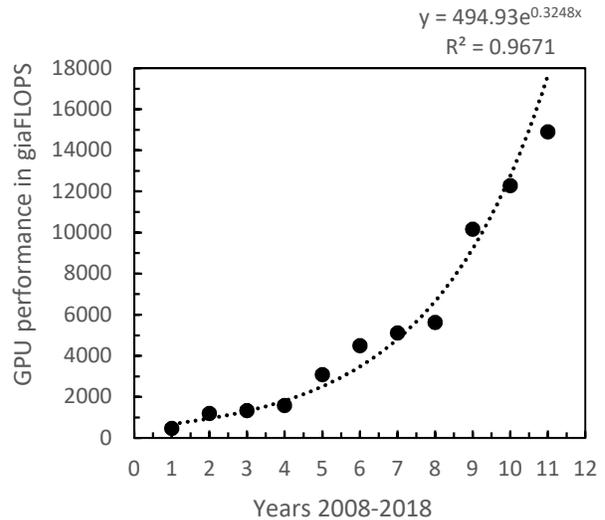

Fig. 10. The performance in gigaFLOPS of GPU devices from NVIDIA since 2008, an exponential regression fits this well allowing prediction of future performance if the same rate of improvement continues.

Fortunately for GPU system users a range of improvements mean the whole software/hardware stack for graphics and AI continues to demonstrate exponential growth [29], as illustrated in Fig. 10. If we use NVIDIA's GPU performance to date as a guide and predict forward, then it should take about twenty years for the petaFLOP performance we have demonstrated in this paper to become the norm on everyday desktop computers. Technologies already announced that will contribute to this include the NVIDIA RTX hardware support for ray tracing [30] and the Blender 2.8 EEVEE algorithms [31].

## 7 CONCLUSIONS

We set out to address three goals:

*Design a new supercomputer application architecture for scalable visualization using the public cloud.* We delivered this using Azure APIs, the krpano viewer and our own task generation and metric monitoring codes.

*Produce the first terapixel urban IoT visualization supporting live daily updates.* We exceeded this target and demonstrated we can render these as frequently as every hour at the same cost as a single node that would take 34 days.

*Undertake a rigorous performance evaluation of cloud supercomputing for visualization applications.* We have demonstrated Gustafson-Barsis' scaling holds for terapixel visualization and achieved twenty sample path traced rendering rates that are equivalent to a real time frame rate of 200Hz for full HD path tracing.

For the future we could use the system to produce daylight correct images of the city hourly throughout the day. We could also use the ray casting power of the system to produce a lightfield rendering which would allow soft focus effects in real time without re-rendering and/or lightfield VR/AR display support.

We don't need more pixels in the current geographic area but a larger digital twin covering the wider suburban area around Newcastle-upon-Tyne would be a benefit for the region. For example, a 33km² map would require a petapixel image that would take 33 days rendering on 1024 nodes, setting a future challenge for our cloud architecture.

Finally, based on our systems experience in this project, we would support the conclusions in [32] that future cloud supercomputing projects would benefit from: improved metrics collection, performance visualization modules, transparent error reporting and simpler purchasing, licensing and cost tracking systems.

## 8 ACKNOWLEDGEMENTS

Many thanks to the following supporters of this research. Northumbria VRV Studio for the VNG 3D model of Newcastle. The National Innovation Centre for Data and our local partners through NICD, Newcastle City Council, for Azure cloud time. Microsoft UK for their invaluable guidance in using the Azure cloud at supercomputer scale. The Alan Turing Institute under the EPSRC grant EP/N510129/1 and for Nicolas Holliman's Turing Fellowship. The EPSRC UKRIC project for funding and supporting the Newcastle Urban Observatory. Siemens UK, in Poole for sponsoring Manu Antony's EPSRC iCASE award.

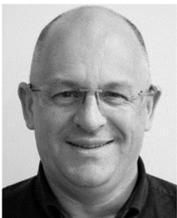

**Nicolas S. Holliman** Ph.D. (University of Leeds 1990) Computer Science, B.Sc. (University of Durham, 1986) joint honours Computing with Electronics. He worked as a software researcher at Lightwork Design Ltd., was Principal Researcher at Sharp Laboratories of Europe Ltd, he was Reader in Computer Science at Durham University and then Professor of Interactive Media at the University of York. He is currently Professor of Visualization at Newcastle University where he heads the Scalable Computing research group and is also a Fellow of the Alan Turing Institute in London. He has published academic articles and patents in visualization, human vision, computer vision, highly parallel computing and autostereoscopic 3D display systems. He is a member of the IS&T, the ACM, a fellow of the Royal Statistical Society and a member of the IEEE Computer Society.

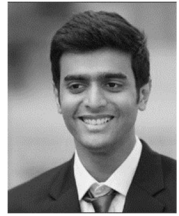

**Manu Antony** B.Eng. Electrical, Electronics and Computer Engineering (Newcastle University 2014). Since 2016 he has been an EPSRC iCASE award PhD student in data visualization sponsored by Siemens Ltd, Poole, UK. He has worked both in the tech industry and the finance industry.

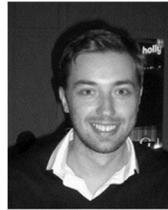

**James Charlton** Ph.D. (Northumbria University, 2011), Architectural Technology B.Sc. (Northumbria University, 2006). He has worked as a Research Assistant, Research Fellow and is now a Senior Lecturer in Architecture at Northumbria University, heading up the VRV Studio and VNG city model. James has published articles in; digital visualisation, virtual city modelling, performance analysis and urban design, he continues to develop his research in these themes.

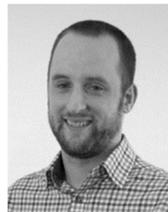

**Stephen Dowsland** M.Sc. (Newcastle University 2009) Town Planning, B.A. Hons (Northumbria University 2007) Geography. He worked in local government town planning before specialising in mapping and front-end web development working at British Airways as part of the major incident planning team. He joined Newcastle University as a Research Software Developer specialising in scalable cloud architectures. He currently works at the National Innovation Centre for Data at Newcastle as a Senior Software Specialist continuing work on scalable cloud and big data processing.

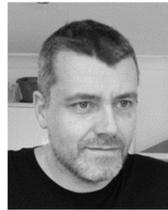

**Phil James** B.A.(Newcastle) is a Senior Lecturer in Engineering. He is currently director of the Newcastle Urban Observatory and co-leads the UK National Observatory Programme. His role is the overall management and direction of the observatory programme and generating strategic partnerships with researchers, civic society and industry. His research is at the intersection of Engineering and Computer Science with a recent focus on IoT and environmental monitoring and how we apply emerging technologies to real-world solutions. He is PI on the EPSRC CORONA (City Observatory Research platfOrm for iNnovation and Analytics) project and participates as Co-I in a research portfolio of interdisciplinary research worth over £15m. He is a Fellow of the Royal Geographical Society.

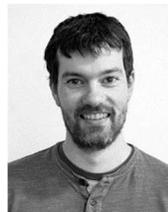

**Mark Turner** M.Sc. (Newcastle University 2012), B.Sc. in Computing (Northumbria University 2008leads the Research Software Engineering team in the Digital Institute at Newcastle. The team focuses on delivering software engineering expertise for research projects across the university. In 2016 he was elected as a trustee for the UK Research Software Engineering Association, contributing to the transformation of the association into a registered charity in 2018. Since joining the university in 2012 he designed and implemented software applications for a number of research projects. Everything from the gamification of stroke rehabilitation physical therapy to mobile applications for alerting stakeholders to damage to rock art carvings and supercomputer scale cloud computing.